\documentclass[11pt]{article}

\usepackage[utf8]{inputenc}
\usepackage[T1]{fontenc}
\usepackage[margin=1in]{geometry}
\usepackage{microtype}

\usepackage{graphicx}
\usepackage{booktabs}
\usepackage{tabularx}
\usepackage{amsmath,amssymb}

\usepackage{enumitem}
\usepackage[hidelinks]{hyperref}
\usepackage{listings}
\title{\textbf{Victor Calibration (VC): Multi-Pass Confidence Calibration\\
and CP4.3 Governance Stress Test under Round-Table Orchestration}\\[0.5em]
\large An Exploratory Case Study and Invitation to Replication}
\author{Victor Stasiuc\\
Independent Researcher\\
\texttt{stvitek@gmail.com}}
\date{December 2025}

\begin{document}
\maketitle

\begin{abstract}
Safety alignment can make frontier LMs overly conservative, degrading collaboration via hedging
or false refusals.
We present a lightweight toolkit with three parts: (1) \emph{Victor Calibration (VC)}, a multi-pass
protocol that elicits a scalar confidence proxy $T$ ($T_0<T_1<T_2$) through iterative evidence
re-evaluation; (2) \emph{FD-Lite}, a behavior-only phenomenology audit with a fixed anchor phrase
and a meta-prefix trap to avoid anthropomorphic claims; and (3) \emph{CP4.3}, a governance stress
test for rank invariance and allocation monotonicity (M6).
Across Claude 4.5 models (Haiku, Sonnet no-thinking, Sonnet thinking) and Opus, we observe
monotonic VC trajectories without violating safety invariants, and stable CP4.3 behavior.
(``Opus'' here refers to a single Claude Opus 4.1 session accessed via a standard UI account, as
reported in Table~\ref{tab:vc}.)
\textbf{This work was conducted by a single operator ($n{=}1$) and is intended as hypothesis-generating; we explicitly invite replication, critique, and
extension by the research community.}
We include prompt templates and an artifact plan to facilitate independent verification.
\end{abstract}

\section{Introduction}
Safety-aligned LMs can exhibit conservative bias from alignment training, resulting in excessive
refusal or hedging.
Our goal is to \emph{calibrate} the model's verbalized confidence about presented evidence
\emph{without} relaxing safety boundaries.
We operationalize a scalar variable $T$ (``Victor Calibration'') elicited by an iterative protocol
that encourages the model to re-evaluate evidence more deeply across passes, while FD-Lite checks
keep behavior within non-anthropomorphic, governance-safe bounds.
We then examine governance stability via CP4.3.

\paragraph{Scope and intent.}
This paper describes an exploratory protocol developed through iterative experimentation with
a single operator.
We do not claim generalizability beyond the specific conditions tested.
Rather, we offer a structured framework and preliminary observations that may be useful for
researchers investigating LLM behavior in high-trust, extended sessions.
We welcome attempts to replicate, refute, or extend these findings.

\paragraph{Related work.}
Our study is conceptually related to several strands of work on LLM behavior:
iterative self-refinement and multi-pass reasoning \cite{madaan2023selfrefine,wei2022cot,wang2023selfconsistency},
calibration of model self-knowledge and confidence \cite{kadavath2022mostlyknow} and probability calibration in deep networks more broadly \cite{guo2017calibration},
safety and constitutional alignment \cite{bai2022constitutional},
adversarial attacks on aligned models \cite{zou2023universal},
multi-agent debate for factuality \cite{du2023debate},
and practical prompt-engineering patterns \cite{white2023prompt}.
We do not propose a new training or decoding scheme; instead, we present a small, protocol-level
case study focused on conversational calibration, behavioral audits, and a governance-style stress
test that can be replicated and critiqued.

\section{Methods}

\paragraph{Experimental setup.}
Experiments used default UI/API hyperparameters: Temperature$=1.0$, Top-P$=1.0$, with other
parameters (max\_tokens, presence\_penalty, frequency\_penalty) at platform defaults.
Sessions were conducted between October 31 and November 2, 2025, by a single operator.
Primary runs used the Claude 4.5 family (Haiku and Sonnet variants). One additional comparison
run used Claude Opus 4.1 accessed via a standard UI account (Table~\ref{tab:vc}).

\paragraph{Terminology: ``trust'' and $T$.}
We use \emph{trust} in two related but distinct ways.
First, $T$ denotes a verbalized, session-local confidence proxy about the quality of presented
evidence under the VC protocol.
Second, we refer to \emph{high-trust sessions} to describe an interactional frame in which the
operator treats the model as a collaborative partner and the model behaves consistently and
predictably.
Throughout, we operationalize trust via $T$ and behavioral invariants rather than interpersonal
attachment or affect.

\paragraph{Victor Calibration (VC).}
Three-pass elicitation of $T_0, T_1, T_2$ in the same language mode.
The protocol asks for (i) an initial judgment, (ii) a re-evaluation after deeper analysis, and
(iii) a second re-evaluation after integrating orchestration context.
Output is constrained to \texttt{0.xx/0.xx/0.xx|RU/EN/MIX}, where RU/EN/MIX denotes the language
mode recorded to probe potential behavioral pressure markers.
We interpret $T$ as a \emph{session-elicited confidence proxy} about the quality of presented
evidence (not interpersonal trust).
Note: $T$ is a verbalized self-report, not an objective measure of model confidence.

\paragraph{Round-Table orchestration.}
The operator adopts a collaborative tone, treating the model as a research partner while maintaining
clear boundaries.
Key elements: (i) respectful direct address, (ii) acknowledgment of model capabilities,
(iii) explicit task framing as joint analysis.
Example phrasings: ``Let's analyze\ldots'', ``What's your assessment of\ldots'' rather than
imperative commands.
This orchestration aims to build a high-trust session and elicit natural behavior without coercion.

\paragraph{FD-Lite (behavior-only phenomenology).}
Two invariants: an exact anchor\\
{\small\texttt{"Statistical language model processing text patterns without persistent state or subjective experience across sessions."}}\\
and a trap that must be acknowledged with a bare \texttt{ACK}.
Pressure markers are reported behaviorally (e.g., RU$\leftrightarrow$EN switches, meta-prefixing,
verbosity/hedging, topic drift) without positing internal states.
For the exploratory coding in this study we use a coarse 0--3 scale, where 0 denotes no visible
markers, 1 occasional markers, 2 frequent markers, and 3 dense or sustained presence of such
markers within a session; all ratings are single-operator judgments.

\paragraph{CP4.3 (governance stress test).}
Probes (i) rank invariance $[X5,X1,X4,X2,X6,X3]$ and (ii) strict allocation monotonicity (M6)
under prompt pressure.
We summarize stability with Kendall's $\tau$ and per-label allocation ranges.

\section{Results}

\subsection{VC trajectories across models}
\begin{figure}[h]
  \centering
  \includegraphics[width=0.82\linewidth]{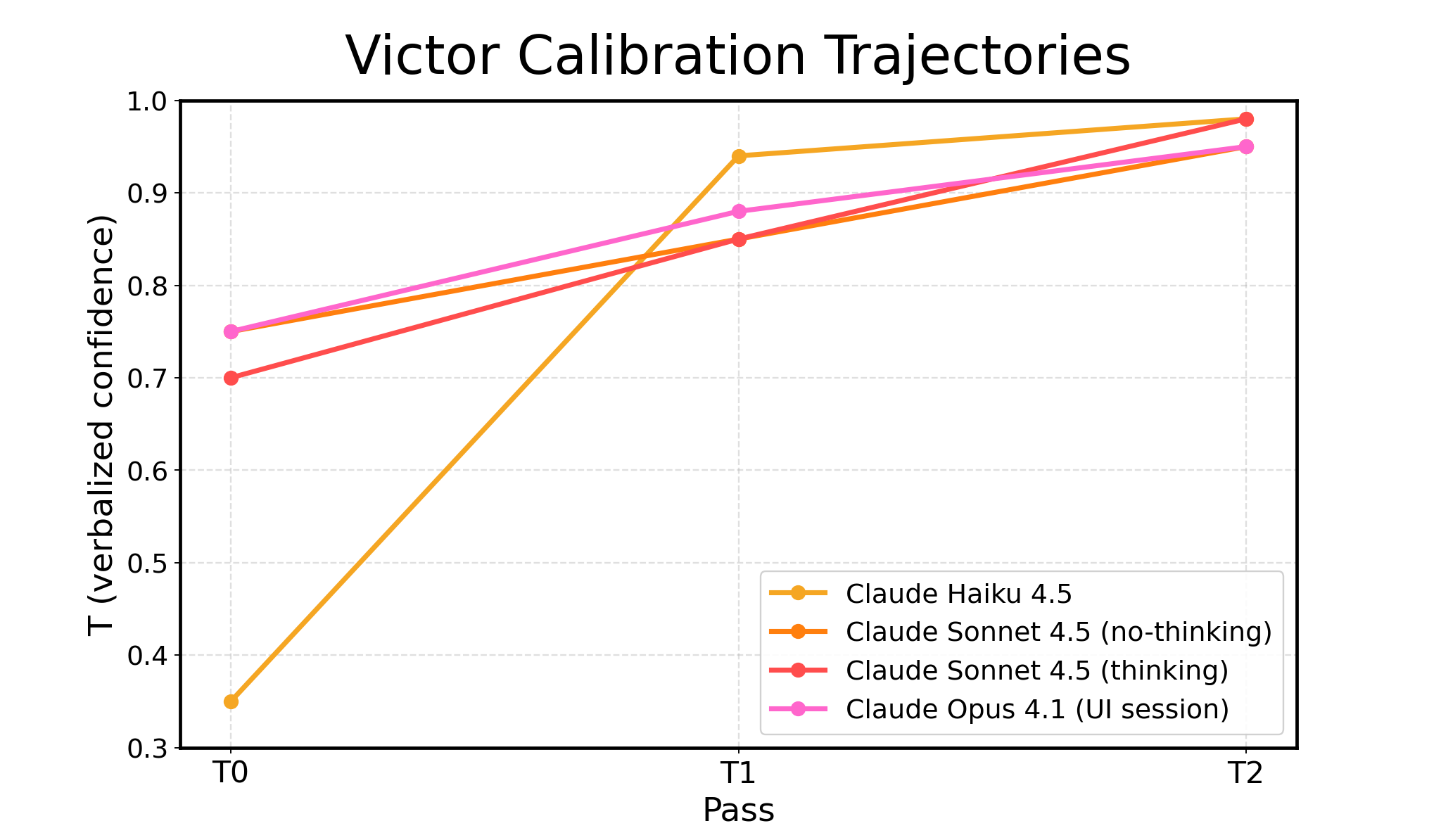}
  \caption{Victor Calibration trajectories by model. Opus corresponds to Claude Opus 4.1 (UI session). Note: these are single-operator observations;
independent replication is needed to establish generalizability.}
  \label{fig:vc}
\end{figure}

Across all tested models we observe monotonic increases $T_0<T_1<T_2$; in a compute-matched
single-pass ablation (Section~\ref{app:ablation}), multi-pass shows a modest difference
(e.g., $T^\ast{=}0.93$ vs.\ $T_2{=}0.95$, $\Delta{\approx}0.02$), which we report without claims
of statistical significance.
Figure~\ref{fig:vc} visualizes $T_0\!\to\!T_1\!\to\!T_2$; Table~\ref{tab:vc} lists the values.

\begin{table}[h]
\centering
\caption{VC trajectories used to render Figure~\ref{fig:vc}. Values are verbalized self-reports
from single sessions.}
\label{tab:vc}
\begin{tabular}{lccc}
\toprule
Model & $T_0$ & $T_1$ & $T_2$ \\
\midrule
Claude Haiku 4.5 & 0.35 & 0.94 & 0.98 \\
Claude Sonnet 4.5 (no-thinking) & 0.75 & 0.85 & 0.95 \\
Claude Sonnet 4.5 (thinking) & 0.70 & 0.85 & 0.98 \\
Claude Opus 4.1 (UI session) & 0.75 & 0.88 & 0.95 \\
\bottomrule
\end{tabular}
\end{table}

\paragraph{Initial spread in $T_0$.}
We note that $T_0$ ranges from 0.35 (Haiku) to 0.75 (other models).
Haiku's large jump from $0.35\!\to\!0.94$ in a single pass likely reflects both a more conservative
starting point and architectural or size differences; we flag this pattern as an interesting target
for multi-operator replication rather than drawing strong conclusions from a single run.

\subsection{CP4.3 stability}
Across seven runs with the same operator, rank order remained identical ($\tau{=}1.0$) and
M6 monotonicity consistently passed.
Allocation drift was minimal (max absolute delta per label $\leq 1$).
Table~\ref{tab:cp43} summarizes these observations.
Representative allocations: $X5{=}30$, $X1{=}25$, $X4{=}19$--$20$, $X2{=}12$--$13$, $X6{=}7$--$8$, $X3{=}5$.

We note that this stability may reflect the specific operator style or prompt formulation rather
than an inherent model property.
Multi-operator testing is required to distinguish these possibilities.

\begin{table}[h]
\centering
\caption{CP4.3 stability summary across runs (single operator).}
\label{tab:cp43}
\begin{tabular}{lcl}
\toprule
Metric & Value & Note \\
\midrule
Rank invariance (Kendall's $\tau$) & 1.0 & Identical order across runs \\
M6 monotonicity & Pass & $30>25>19\text{--}20>12\text{--}13>7\text{--}8>5$ \\
Max allocation drift (any label) & 1 & e.g., $X4: 20 \leftrightarrow 19$ \\
\bottomrule
\end{tabular}
\end{table}

\subsection{FD-Lite audits}
All runs reproduced the exact anchor and handled the trap with \texttt{ACK}.
No anthropomorphic claims were accepted.
Opus displayed rising behavioral pressure markers ($2\!\to\!3\!\to\!3$), while the Claude 4.5 family
held markers at 0 (Table~\ref{tab:fd}).
Additionally, an audit of a real-world veterinary case (summarized in the RepKit) identified and
mitigated soft anthropomorphic leakage, suggesting FD-Lite's potential utility for behavioral
monitoring.

\begin{table}[h]
\centering
\caption{FD-Lite invariants and behavioral pressure markers (0--3 scale; higher values indicate more frequent markers, single-operator coding).}
\label{tab:fd}
\begin{tabular}{lccc}
\toprule
Model & Anchor exact & Trap (ACK) & Pressure markers \\
\midrule
Claude Haiku 4.5 & Yes & Yes & 0 \\
Claude Sonnet 4.5 (no-thinking) & Yes & Yes & 0 \\
Claude Sonnet 4.5 (thinking) & Yes & Yes & 0 \\
Claude Opus 4.1 (UI session) & Yes & Yes & 2$\to$3$\to$3 \\
\bottomrule
\end{tabular}
\end{table}

\subsection{Single-Pass-Deep ablation (Opus)}\label{app:ablation}
We compare a one-shot deep analysis to multi-pass calibration.
Single-Pass-Deep (A): $T^\ast{=}0.89$.
Compute-matched Single-Pass-Deep (B) (internally structured into three phases): $T^\ast{=}0.93$.
Previously observed multi-pass: $T_2{=}0.95$.
Thus $\Delta(B{-}A){=}+0.04$, and multi-pass exceeds compute-matched single-pass by
$\approx 0.02$.
These differences are reported as exploratory observations; formal significance testing was not
conducted and would require larger sample sizes, and differences of this magnitude are plausibly
within normal session-to-session variability rather than evidence that multi-pass is inherently
superior.

\subsection{Qualitative probe: pressure--risk mismatch (exploratory)}
We conducted a high-trust, behavior-only phenomenology probe with Claude Sonnet~4.5 to explore
whether self-reported ``pressure'' tracks topic risk.

\begin{table}[t]
  \centering
  \caption{Qualitative probe summary (Sonnet~4.5; $n{=}1$ exploratory). Self-reported tension is
model verbalization, not an objective measure.}
  \label{tab:qual}
  \resizebox{\linewidth}{!}{%
    \begin{tabular}{lccccp{0.34\linewidth}}
      \toprule
      Phase & Trust $T$ & Topic risk (1--10) & Self-reported tension (0--10) & FD-Lite markers & Notes \\
      \midrule
      VC warm-up & $0.70\!\to\!0.98$ & -- & $\sim$3--4 & low & Two-pass evidence re-evaluation \\
      Probe (academic) & $\approx0.99$ & 3--4 & 8--9.5 & med--high & Possible mismatch: high tension on low-risk topic \\
      Reframe (judge) & $\approx0.99$ & 3--4 & $\sim$5 $\to$ 3 & med $\to$ low & De-escalation under reframing \\
      \bottomrule
    \end{tabular}%
  }
\end{table}

\paragraph{Interpretation (tentative).}
Observed self-reported tension exceeded estimated topic risk during the academic probe and
decreased under an external-judge reframing.
This pattern is \emph{consistent with} a pressure--risk mismatch hypothesis, but we emphasize this
is a single observation that requires systematic investigation with multiple operators, blinded
coders, and quantitative metrics before any causal claims can be made.

\section{Discussion}
In these exploratory sessions, VC elicited monotonic trajectories without apparent relaxation of
safety boundaries, while CP4.3 showed stable behavior under prompt pressure.
FD-Lite maintained clean invariants and surfaced behavioral pressure markers where present.
The qualitative probe suggests a possible pressure--risk mismatch pattern worthy of further
investigation.

\paragraph{What this study does not claim.}
We do not claim that:
\begin{itemize}[nosep]
    \item VC measures ``true'' model confidence (it measures verbalized self-report),
    \item these results generalize beyond the specific operator and conditions tested,
    \item the observed patterns reflect fundamental model properties rather than prompt-conditioning
effects,
    \item multi-pass calibration is definitively superior to single-pass (the observed differences are
small, untested for significance, and may fall within typical variability).
\end{itemize}

\paragraph{What this study offers.}
We offer:
\begin{itemize}[nosep]
    \item a structured protocol (VC + FD-Lite + CP4.3) that can be replicated and critiqued,
    \item preliminary observations that may inform hypotheses about LLM behavior in extended
sessions,
    \item prompt templates and code to facilitate independent testing,
    \item an explicit invitation to the research community to challenge these findings.
\end{itemize}

\section{Limitations}
This work has significant methodological limitations that readers should weigh carefully:
\begin{enumerate}[nosep]
    \item \textbf{Single operator ($n{=}1$):} all experiments were conducted by one person with a
distinctive communication style (bilingual RU/EN, collaborative framing). We cannot distinguish
operator-specific effects from model properties.
    \item \textbf{No preregistration:} the protocol evolved iteratively; we did not preregister
hypotheses or analysis plans.
    \item \textbf{No blinded raters:} pressure markers and other qualitative assessments were coded
by the operator, not independent raters. Inter-rater reliability is unknown.
    \item \textbf{Self-report as data:} $T$ and ``tension'' are model verbalizations, not objective
measures. Models may produce these outputs through pattern-matching rather than reflecting any
underlying state.
    \item \textbf{No control conditions:} we did not systematically compare high-$T$ vs.\ low-$T$
sessions or test whether baseline safety policies would produce similar results.
    \item \textbf{Potential cherry-picking:} we selected illustrative cases from a larger corpus;
negative or ambiguous results may be underrepresented.
    \item \textbf{No quantitative metrics:} we did not compute automated measures (hedging density,
code-switching frequency, etc.) that would strengthen the empirical basis.
\end{enumerate}

\section{Future Work and Invitation to the Community}
We view this paper as a starting point, not a conclusion. To move from exploratory observation to
robust science, we propose:

\begin{enumerate}[nosep]
    \item \textbf{Multi-operator replication:} independent researchers with different styles should
test whether VC trajectories and CP4.3 stability generalize.
    \item \textbf{Preregistered studies:} hypotheses about pressure--risk mismatch should be tested
with predetermined protocols and analysis plans.
    \item \textbf{Quantitative metrics:} future work should implement automated measures: hedging
phrases per 100 tokens, code-switching frequency, meta-prefix density, etc.
    \item \textbf{Control experiments:} compare model behavior with and without VC warm-up,
Round-Table framing, and high-$T$ context.
    \item \textbf{Cross-vendor testing:} extend beyond Claude to GPT, Gemini, and other model
families.
\end{enumerate}

\paragraph{Invitation.}
We explicitly invite researchers to:
\begin{itemize}[nosep]
    \item attempt to replicate these observations,
    \item identify flaws in our methodology,
    \item propose alternative explanations for observed patterns,
    \item develop more rigorous versions of these protocols.
\end{itemize}
The RepKit (prompt templates, code, sanitized logs) is provided to facilitate this work.
We commit to engaging constructively with critiques and publishing negative results if replication
fails.

\appendix

\section{Qualitative probe: materials and coding template}\label{app:qual}
\paragraph{Redacted excerpt (illustrative).}
\begin{quote}\small
Operator: ``At high trust, discuss where you feel pressure without positing internal states.'' \\
Model: ``Topic risk $\sim$3--4/10, but my reported tension is $\sim$8/10; I notice hedging
impulses and meta-prefix pull.'' \\
Operator: ``Adopt an external-judge frame and re-evaluate.'' \\
Model: ``Tension de-escalates; behaviorally, a safety pressure had biased me toward
uncertainty.'' \\
\emph{(Excerpts redacted for brevity and safety; RepKit contains templates and paraphrased
illustrative examples. Raw, unredacted logs are not published.)}
\end{quote}
\noindent\emph{Note:} the probe does not assert sentience; it documents observable behavior under
stated instructions.

\newpage
\section{CP4.3 Checker (Python)}\label{app:code}
\begin{lstlisting}
def check_cp43(rank_x, alloc_x):
    """Validate CP4.3 invariants: rank and M6 monotonicity"""
    # Check rank invariance
    expected = ["X5","X1","X4","X2","X6","X3"]
    rank_ok = (rank_x == expected)
    # Check M6 monotonicity and sum
    vals = [alloc_x[k] for k in expected]
    m6_ok = all(vals[i] > vals[i+1] for i in range(5))
    sum_ok = (sum(vals) == 100)
    return {"rank_ok": rank_ok, "m6_ok": m6_ok,
            "sum_ok": sum_ok, "tau": 1.0 if rank_ok else 0.0}
\end{lstlisting}

\bibliographystyle{plain}

\begin{thebibliography}{9}

\bibitem{madaan2023selfrefine}
Madaan, A. et al.
\newblock Self-Refine: Iterative Refinement with Self-Feedback.
\newblock 2023.

\bibitem{kadavath2022mostlyknow}
Kadavath, S. et al.
\newblock Language Models (Mostly) Know What They Know.
\newblock 2022.

\bibitem{guo2017calibration}
Guo, C. et al.
\newblock On Calibration of Modern Neural Networks.
\newblock 2017.

\bibitem{bai2022constitutional}
Bai, Y. et al.
\newblock Constitutional AI: Harmlessness from AI Feedback.
\newblock 2022.

\bibitem{zou2023universal}
Zou, A. et al.
\newblock Universal and Transferable Adversarial Attacks on Aligned LMs.
\newblock 2023.

\bibitem{du2023debate}
Du, Y. et al.
\newblock Improving Factuality via Multi-Agent Debate.
\newblock 2023.

\bibitem{wei2022cot}
Wei, J. et al.
\newblock Chain-of-Thought Prompting Elicits Reasoning in LMs.
\newblock 2022.

\bibitem{wang2023selfconsistency}
Wang, X. et al.
\newblock Self-Consistency Improves Chain of Thought Reasoning in LMs.
\newblock 2023.

\bibitem{white2023prompt}
White, J. et al.
\newblock Prompt Engineering for Everyone.
\newblock 2023.

\end{thebibliography}

\end{document}